\documentstyle[12pt,draft]{article}

\def\v#1{\mbox{\boldmath$#1$}}

\begin{document}

\vspace{2cm}

\title{The role of the $\Delta$(1232) on the longitudinal response\\
in the inclusive electron scattering reaction}

\author{E. Bauer\thanks{Fellow of the Consejo Nacional
de Investigaciones Cient\'{\i}ficas y T\'ecticas, CONICET.} \\
Departamento de F\'{\i}sica, Facultad de Ciencias Exactas,\\
Universidad Nacional de La Plata, La Plata, 1900, Argentina.}
\maketitle

\vspace{1cm}

\begin{abstract}
We have performed a many-body calculation of the longitudinal nuclear $(e,e')$
reaction employing a Second RPA (SRPA)
formalism which contains the $\Delta$(1232).
More explicitly, our scheme contains RPA correlations
as well as Hartree-Fock and second order self-energies, where
an accurate evaluation of exchange terms is achieved.
Using this formalism we have evaluated the longitudinal
response function for $^{40}Ca$.
We give final results at momentum transfers ranging from
300 up to 500 MeV/c, obtaining a good agreement with data.
\end{abstract}

\vspace{1.3cm}


\vspace{.5cm}


\newpage
\section{INTRODUCTION}
The quasi-elastic peak in inclusive electron scattering
for medium and heavy nuclei
has been paid much attention in recent years.
The good description
of the cross section in the former work of Monitz {\it
et al.}~\cite{kn:mo71} was suddenly broken when
the longitudinal and transverse responses were
experimentally separated by Meziani {\it et al.}~\cite{kn:me84}.
Even though the total cross section is
well reproduced by a Fermi gas, this simple
model overestimate data in the longitudinal
channel, while it does the opposite in the
transverse one. More recently Williamson {\it et
al.}~\cite{kn:wi97} have presented new
measurements with an increase (decrease) of the
longitudinal (transverse) nuclear response. Still,
the discrepancy remains.

Many theoretical efforts have been developed to
overcome this problem.
Let us resume the main ones. One
group ascribes the problem to modifications of the
nucleon properties within the nuclear medium by the
employment of small effective masses, swelling of
nucleons, abnormal nucleon form factors, etc. (see
Ref.~\cite{kn:no81}). Even though this approach is successful
for the longitudinal channel, it fails in the
transverse one. The other group, to which belongs
the present work, is based on the many-body
theory.

Many-body calculations were performed in finite
nucleus \cite{kn:ca84}-\cite{kn:sl95} or in nuclear matter
\cite{kn:al84}-\cite{kn:am99b}.
Fortunately, surface effects for heavy and medium
nuclei in the energy-momentum region of interest are not
very important and the nuclear matter formalism is
a good approximation for the nuclear system, once
a variable Fermi momentum or the local density
approximation is used. On the other side, the
many-body problem is very complex: initial and
final state interactions, meson exchange currents,
etc., should be considered, which implies a huge
numerical effort, even in nuclear matter.

Perhaps the simplest way of introducing nuclear
correlations is the RPA (see Refs. \cite{kn:de85},
\cite{kn:sh89}, \cite{kn:ba96} and \cite{kn:ba98}). In the same line, a step
further in difficulty is the SRPA~\cite{kn:co88}, \cite{kn:ba00},
which opens decay channels beyond one~particle-one~hole ($1p1h$)
states and the Extended RPA (ERPA)
(\cite{kn:ta88}, \cite{kn:ba95} and \cite{kn:ba98b}), adding
more complex ground state correlations than in
the RPA. Others approaches are
the Correlated Basis Function (CBF)~\cite{kn:fa87},
the Green function method \cite{kn:ch89} and the one
Boson Loop Expansion (BLE)~\cite{kn:ce97}, \cite{kn:am99b}. Looking at these formalism in
terms of Feyman diagrams, all these approaches
should converge to the same result, once the
proper number of diagrams are considered. In spite of
the sophistication of all these models,
a clear understanding of the nuclear response at any
momentum transfer is still elusive.

In some recent works the role of the $\Delta$(1232) on the longitudinal
channel was explored in some recent works.
Gil {\it et al.}~\cite{kn:gi97} study both channels within
an elaborate formalism which includes two-body corrections to the excitation
operator, initial and final state interaction and the $\Delta$.
The $\Delta$ was considered by modifications in the external
operator, self-energy insertions and in the nuclear
interaction by means of the ring series with the
$\Delta$ (polarization effect). The $\Delta$ is also
incorporated in the longitudinal response in the works of
Cenni {\it et al.}~\cite{kn:ce97} and
Amore {\it et al.}~\cite{kn:am99b} where the Boson Loop Expansion is employed.
In Amaro {\it et al.}~\cite{kn:am99} (and also in Ref.~\cite{kn:am99b}),  it
is addressed that a $\Delta - hole$ pair can be
directly excited in the longitudinal channel.
In Ref.~\cite{kn:ba98} we have also examined both longitudinal and transverse
responses with the explicit inclusion of the $\Delta$. In that
work, however, we concentrate on the transverse channel, with
special attention on the RPA-exchange terms, concluding that an
accurate evaluation of them is important. In that early work, the
$\Delta$ has entered into the longitudinal response only through
the second order self-energy, as there were no RPA terms in the
longitudinal channel due to the particular election of the
residual interaction.

In the present work we have developed
a SRPA formalism which employs a more elaborate residual
interaction than in our previous work~\cite{kn:ba98}.
The SRPA contains the RPA correlations together with the Hartree-Fock
and second order self-energies. Exchange terms were
carefully taken into account and the $\Delta$ was incorporated both in the
many-body calculation and by it direct excitation.

The work is organized as follows. In Section II, we
present the formalism. In Section III we analyze the effect of
each ingredient in our model and analyze results for $^{40}Ca$.
Finally, in Section IV some conclusions are given.

\newpage

\section{FORMALISM}
Even though the main
ingredients of our formalism were already presented in
Refs. \cite{kn:ba98} and \cite{kn:ba00},
here we have preferred to show a complete
version of it. This is because there are some elements
which are particular to the
nuclear longitudinal response with the $\Delta$ which
should be discussed.

The nuclear response function to an electromagnetic  probe
in the longitudinal channel is defined as,
\begin{equation}
R_L(\v{q},\omega) = -\frac{1}{\pi} \,
Im \, <|{\cal O}^{\dag}_L G(\omega) {\cal O}_L |>
\label{eq:rf1},
\end{equation}
where $\v{q}$ represents the magnitude of the
three momentum transfer by the electromagnetic probe ${\cal O}_L$, $\omega$ is
the excitation energy and $|>$ is the uncorrelated nuclear ground
state. Ground state
correlations beyond RPA are not analyzed in this work. The
polarization propagator is given by,
\begin{equation}
G(\omega) = \frac{1}{\omega - H + i \eta}\ -
\frac{1}{\omega + H - i \eta}\
\label{eq:green},
\end{equation}
where $H$ is the nuclear Hamiltonian. As usual, $H$ is
separated into a one-body part, $H_0$, and a residual
interaction $V$.

We present now two projection operators $P$ and $Q$. The
action of $P$ is to project into the ground state, the one
particle-one hole ($ph$) and one delta-one hole ($\Delta h$)
configurations. While $Q$ projects into the residual
$n_p$ particle-$n_h$ hole-$n_{\Delta}$ delta configurations.
More explicitly,
\begin{equation}
P = |><| + P_N + P_{\Delta}
\label{eq:prop},
\end{equation}
with
\begin{equation}
P_N = \sum_{1p1h} |1p1h><1p1h|
\label{eq:propn},
\end{equation}
\begin{equation}
P_{\Delta} = \sum_{1 \Delta 1h} |1 \Delta 1h><1 \Delta 1h|
\label{eq:propd}
\end{equation}
and
\begin{equation}
Q  =  \sum_{ \begin{array}{c}
     n_h \geq 2 \\ 0 \leq n_p \leq n_h
       \end{array} }
|n_p p \; (n_h-n_p) \Delta \; n_h h><n_p p \; (n_h-n_p) \Delta \; n_h h|
\label{eq:proq},
\end{equation}
where we have introduced $P_N$ and $P_{\Delta}$ for convenience.
It is easy to
verify that $P + Q = 1$, $P^2 = P$,
$Q^2 = Q$, and $PQ = QP = 0$ and also,
$P_i P_j = \delta_{i j} P_i$ and $P_i Q = Q P_i = 0$ ($i = N, \Delta$).

We insert now the identity into eq.~(\ref{eq:rf1}). Note
that the external operator can connect the uncorrelated
ground state only to the $P$ space if we use a one body operator for ${\cal O}_L$, we have,
\begin{equation}
R_L(\v{q},\omega) = -\frac{1}{\pi} \ Im
<|{\cal O}_L^{\dag} \; P \; G_{PP}(\omega) \; P \; {\cal O}_L |>
\label{eq:rn2},
\end{equation}
where $G_{PP} \equiv P G P$. It is easy to see that,
\begin{equation}
G_{PP}(\omega) = \frac{1}{\omega - H_{PP} -
\Sigma^{PQP} + i \eta}\ -
\frac{1}{\omega + H_{PP} +
\Sigma^{PQP} - i \eta}\
\label{eq:gw2},
\end{equation}
where,
\begin{equation}
\Sigma^{PQP} = V_{PQ} \frac{1}{ \omega -H_{QQ} + i \eta}\ V_{QP} -
       V_{PQ} \frac{1}{ \omega +H_{QQ} - i \eta}\ V_{QP}
\label{eq:self},
\end{equation}
with obvious definitions for $H_{PP}$, etc.

Until now, the formulation is quite
general. The next step is to adopt a model for the external operator,
the nuclear Hamiltonian and the nuclear states. The nuclear interaction
is discussed in the next section, while the nuclear structure
problem is the main issue of this section. Before this, we
show a model for ${\cal O}_L$, which
is divided into two contributions,
\begin{equation}
{\cal O}_L = {\cal O}_{ph} + {\cal O}_{\Delta h},
\label{eq:opnd}
\end{equation}
where the action of ${\cal O}_{ph}$ (${\cal O}_{\Delta h}$) is to
create a $ph$ ($\Delta h$) pair.
The matrix elements for ${\cal O}_{ph}$ and ${\cal O}_{\Delta h}$
were taken from Refs. \cite{kn:am96} and \cite{kn:am99}, respectively,
\begin{equation}
<1p1h|{\cal O}_{ph}|> = G_E^p(\v{q},\omega) \frac{\kappa}{\sqrt{\tau}}
{\chi}^{\dag}_{s_p} {\chi}^{\dag}_{t_p} \frac{1+\tau_3}{2}
{\chi}_{s_h} {\chi}_{t_h}
\label{eq:opn}
\end{equation}
and
\begin{equation}
<1 \Delta 1h|{\cal O}_{\Delta h}|> = G_{\Delta}(\v{q},\omega) \frac{\sqrt{1+\tau'}}{m^2}
{\chi}^{\dag}_{s_\Delta} {\chi}^{\dag}_{t_\Delta} \v{S}^{\dag}
\cdot (\v{h} \times \v{q}) \; T_3 \;
{\chi}_{s_h} {\chi}_{t_h}
\label{eq:opd}
\end{equation}
where the $s$'s and the $t$'s stand for
the spin and isospin quantum numbers,
$\v{h}$ is the momentum carried by the hole, $\kappa=q/(2m)$,
$\tau=(q^2-\omega^2)/(4 m^2)$,
$\tau'=\frac{m}{m_\Delta} \, (\tau+(\frac{m_\Delta-m}{2 m_\Delta})^2)$,
and $m$ ($m_\Delta$)is the nucleonic ($\Delta$) mass.
In eq.~(\ref{eq:opd}) the Pauli matrices $\v{\sigma}$
and $\tau_3$ were replaced by the corresponding transitions
matrices $\v{S}$ and $T_3$ \cite{kn:br75}. The
electromagnetic form factors are,
\begin{equation}
G_E^p(\v{q},\omega) = \frac{1}{(1+4.97 \tau)^2}
\label{eq:gep}
\end{equation}
and
\begin{equation}
G_{\Delta}(\v{q},\omega) = 2.97 \; G_E^p(\v{q},\omega) \; (1-\frac{\omega^2-q^2}{3.5
(GeV/c)^2})^{-1/2}.
\label{eq:gdelt}
\end{equation}
In the longitudinal channel there is no interference term between
${\cal O}_{ph}$ and ${\cal O}_{\Delta h}$.
This is at variance with the case of the transversal one and it is a consequence
of the presence of the hole momentum in ${\cal O}_{\Delta h}$.
This property allows to rewrite Eq.~(\ref{eq:rn2}) into two terms,
\begin{equation}
R_L = R_{ph} + R_{\Delta h},
\label{eq:rn3}
\end{equation}
where
\begin{equation}
R_{ph} = -\frac{1}{\pi} \ Im
<|{\cal O}_{ph}^{\dag} \; P_N \; G_{P_N P_N}(\omega) \; P_N \; {\cal O}_{ph}|>
\label{eq:rnn}
\end{equation}
and
\begin{equation}
R_{\Delta h} = -\frac{1}{\pi} \ Im <|{\cal O}_{\Delta h}^{\dag} \; P_\Delta \;
G_{P_\Delta P_\Delta}(\omega) \; P_\Delta \; {\cal O}_{\Delta h}|>.
\label{eq:rnd}
\end{equation}

In the next two subsections we analyze separately
$R_{ph}$ and $R_{\Delta h}$.

\subsection{$R_{ph}$ contribution}

The starting point of this subsection is Eq.~(\ref{eq:rnn}) where
our concern is to understand this expression in terms of it
physical ingredients. To start with,
we replace $P_N$ by its explicit expression,
\begin{eqnarray}
R_{ph} & = & -\frac{1}{\pi} \ Im \sum_{1p1h, \, 1p'1h'}
<|{\cal O}_{ph}^{\dag}|1p1h><1p1h|
\frac{1}{\omega - H_{P_N P_N} - \Sigma^{P_N Q P_N} + i \eta}
 |1p'1h'> \nonumber \\
 & & \times <1p'1h'| {\cal O}_{ph}|>
\label{eq:rph}
\end{eqnarray}
where for simplicity we have shown only the forward going
contribution.
We consider now the energy denominator of Eq. (\ref{eq:rph}). For
$H_{P_N P_N}$ we have,
\begin{equation}
<1p1h| H_{P_N P_N} |1p'1h'> = \delta_{ph,p'h'} (E_{1p1h}^{(0)} + \Sigma_{1p1h}^{HF})
+ V_{1p1h,1p'1h'}
\label{eq:edc},
\end{equation}
where $E_{1p1h}^{(0)}$ is the kinetic energy of the particle-hole
pair, $\Sigma_{1p1h}^{HF}$ is the Hartree-Fock self-energy contribution
and $V_{1p1h,1p'1h'}$ is the residual interaction
between two particle-holes pairs. More explicitly,
we can define the energy of a $1p1h$ pair up to first order
in the residual interaction as,
\begin{equation}
E_{1p1h}^{(1)} \equiv E_{1p1h}^{(0)} + \Sigma_{1p1h}^{HF} =
\varepsilon^{(1)}(p) - \varepsilon^{(1)}(h)
\label{eq:ham2c},
\end{equation}
where in addition, we have introduced the energy of a single
nucleon with momentum $\v{p}$ as,
\begin{equation}
\varepsilon^{(1)}(p) \equiv \frac{p^2}{2 m} \ +
\Sigma^{HF}(p).
\label{eq:ham2b}
\end{equation}
The analytical expression for $\Sigma^{HF}(p)$ depends on the
interaction employed and it is discussed in the next section.

The action of the $\Sigma^{P_N Q P_N}$
self-energy is to connect the $P_N$-space with the $Q$ one.
This means
that $\Sigma^{P_N Q P_N}$ opens the decay channel of a $1p1h$ pair into
a $2p \, 2h$ configuration, a $1 \Delta 1p 2h$ one and also more complex
configurations like $2 \Delta 2h$.

The matrix elements of $\Sigma^{P_N Q P_N}$ are written as,
\begin{equation}
<1p1h| \Sigma^{P_N Q P_N} |1p'1h'>
\cong \; \delta_{ph,p'h'} \Sigma^{P_N Q P_N}_{1p1h}(\v{h},\v{q},\omega)
\label{eq:sig1}
\end{equation}
where we have kept only the diagonal terms of the self energy.
This approximation is based on numerical reasons
already discussed in~Ref. \cite{kn:ba98}. Also, in this work
we have studied in detail $\Sigma^{P_N Q P_N}_{1p1h}$
splitting it into four components, depending on the character of
the intermediate configuration $Q$, as,
\begin{equation}
\Sigma^{P_N Q P_N}_{1p1h}(\v{h},\v{q},\omega) = \sum_{i=1}^{4}
\Sigma^{P_N Q_i P_N}_{1p1h}(\v{h},\v{q},\omega)
\label{eq:senn}.
\end{equation}
where $Q_1$ represents a $1p1h$ bubble attached to a $1p1h$,
$Q_2$ is a $1 \Delta 1h$ bubble attached to a $1p1h$,
$Q_3$ is a $1p1h$ bubble attached to a $1 \Delta 1h$ and
finally, $Q_4$ is a $1\Delta 1h$ bubble attached to a $1\Delta 1h$.
Diagrams of each of the $Q_i$ can be found in Fig.~6 of
Ref.~\cite{kn:ba98}. Their analytical
expressions are also found in the same reference.

Also through the numerical analysis it turns out that
the dependence of the self-energy over the
hole momentum is not very strong. This allows us to
make an average over it,
\begin{equation}
\Sigma^{P_N Q P_N}_{1p1h}(\v{h},\v{q},\omega) \; \cong \;
\Sigma^{P_N Q_i P_N}_{1p1h}(\v{q},\omega)
\label{eq:sigap}
\end{equation}
where
\begin{equation}
\Sigma^{P_N Q_i P_N}_{1p1h}(\v{q},\omega) \equiv
\frac{1}{ \frac{4}{3} \ \pi} \ \int d^3 h \ \, \theta(k_F-|\v{h}|) \;
\Sigma^{P_N Q_i P_N}_{1p1h}(\v{h},\v{q},\omega)
\end{equation}
and $k_F$ is the Fermi momentum.
Now Eq. (\ref{eq:rph}) reads,
\begin{eqnarray}
R_{ph} & = & -\frac{1}{\pi} \ Im \sum_{1p1h, \, 1p'1h'}
<|{\cal O}_{ph}^{\dag}|1p1h><1p1h|
\frac{1}{\omega - \delta_{ph,p'h'} \; E_{1p1h}^{(2)} - V_{1p1h,1p'1h'}+ i \eta}
\nonumber \\
& & \times
|1p'1h'> <1p'1h'| {\cal O}_{ph}|>
\label{eq:rph2}
\end{eqnarray}
where
\begin{equation}
E_{1p1h}^{(2)} =  \, E_{1p1h}^{(0)} + \Sigma_{1p1h}^{HF} \, + \,
\Sigma^{P_N Q P_N}_{1p1h}(\v{q},\omega)
\label{eq:spp4b}.
\end{equation}
The non-diagonal term of the interaction $V_{1p1h,1p'1h'}$,
is what makes Eq.~(\ref{eq:rph2}) hard to evaluate. Due to this
reason, it is convenient
to show first the result without $V_{1p1h,1p'1h'}$. We define,
\begin{equation}
R_{ph}^{(i)} \equiv -\frac{1}{\pi} \ Im \sum_{1p1h}
|<1p1h| {\cal O}_{ph}|>|^2 \;
\frac{1}{\omega -  E_{1p1h}^{(i)} + i \eta}
\label{eq:rphi}
\end{equation}
where the $1p1h$ energies $E_{1p1h}^{(i)}$ with $i=$ 0, 1 and 2, were
already defined in Eqs.~(\ref{eq:edc}, \ref{eq:ham2b}) and (\ref{eq:spp4b}),
respectively. For $i=0$ we obtain the free response,
\begin{equation}
R_{ph}^{(0)}(\v{q},\omega) = \sum_{1p1h} \; |<1p1h| {\cal O}_{ph}|>|^2 \;
\delta (\omega -  (\frac{(\v{h} + \v{q})^2}{2 m_p} - \frac{\v{h}^2}{2 m_h}) )
\label{eq:rph0}
\end{equation}
where $\v{h} + \v{q}$ is the particle momentum  and we have
allowed the particles and the holes to have different masses.
Performing the summation over spin and isospin and
making the conversion of sums over momenta into integrals we have,
\begin{equation}
R_{ph}^{(0)}(\v{q},\omega) = \frac{3 \pi^2 A}{k_F^3} \; (G_E^p(\v{q},\omega))^2 \;
\frac{\kappa^2}{\tau}
\; {\cal L}(\v{q},\omega,m_p,m_h)
\label{eq:rph0l}
\end{equation}
where $A$ is the mass number and ${\cal L}(\v{q},\omega,m_p,m_h)$
is a modified Lindhard
function defined in Appendix A. The Goldstone diagram
of $R_{ph}^{(0)}$ is drawn as the first one in the {\it r.h.s.} of
Fig.~1.

The response function for $i=1$ is,
\begin{equation}
R_{ph}^{(1)}(\v{q},\omega) = \sum_{1p1h} \; |<1p1h| {\cal O}_{ph}|>|^2 \;
\delta (\omega -  (\varepsilon^{(1)}(|\v{h} + \v{q}|) - \varepsilon^{(1)}(h)) )
\label{eq:rph1}
\end{equation}
The expression within the $\delta$-function is a transcendental
equation due to the presence of $\Sigma_{ph}^{HF}$ and is solved
numerically. Alternatively,
in Ref. \cite{kn:br96}
it was discussed that the $R_{ph}^{(1)}$ response is well
reproduced by $R_{ph}^{(0)}$ but using two effective masses (one
for particles and one for holes), and an energy shift. Adopting
this approximation, we can write,
\begin{equation}
R_{ph}^{(1)}(\v{q},\omega) = \frac{3 \pi^2 A}{k_F^3} \; (G_E^p(\v{q},\omega))^2 \;
\frac{\kappa^2}{\tau}
\; {\cal L}(\v{q},{\bar \omega},m_p,m_h)
\label{eq:rphlm}
\end{equation}
where ${\bar \omega} = \omega + \Delta \omega$ and the values for
$m_p$, $m_h$ and $\Delta \omega$ depend on the interaction and on the
momentum transfer. The first three diagrams contributing to $R_{ph}^{(1)}$
are the first three ones of Fig.~1. The fact that the Hartree-Fock
self-energy is in the energy denominator implies that this contribution is sum
up to infinite order.

Finally we consider $R_{ph}^{(2)}$. At variance with the case of
$R_{ph}^{(1)}$ where the the Hartree-Fock self-energy is a pure real
function of the particle (or hole) momentum, the $\Sigma^{P_N Q
P_N}_{ph}$ self-energy has a real and an imaginary part. It is
easy to see that $R_{ph}^{(2)}$ can be written as,
\begin{equation}
R_{ph}^{(2)}(\v{q},\omega) = - \, \frac{1}{\pi} \ \int_{0}^{\infty} dE
\; R_{ph}^{(1)}(\v{q},E)
\frac{Im \Sigma^{P_N Q P_N}_{1p1h}(\v{q},\omega)}{(E - \omega +
Re \Sigma^{P_N Q P_N}_{1p1h}(\v{q},\omega))^2 +
(Im \Sigma^{P_N Q P_N}_{1p1h}(\v{q},\omega))^2 } \
\label{eq:rphse2},
\end{equation}
The lower order contributions which comes entirely from
$\Sigma^{P_N Q_i P_N}_{1p1h}$, are shown in the second
line of Fig.~1 (in upward order of $i$).

We turn now to the discussion of $R_{ph}$ with the inclusion
of the $V_{1p1h,1p'1h'}$ term.
We write down from Eq.~(\ref{eq:rph2}) the matrix element of the
polarization propagator,
\begin{equation}
G(\omega)_{1p1h, \, 1p'1h'} =
\frac{1}{\omega - \delta_{ph,p'h'} \; E_{1p1h}^{(2)} - V_{1p1h,1p'1h'}+ i \eta}
\label{eq:greenph}
\end{equation}
where as mentioned above, we show for simplicity only the forward
going contribution. To treat this, the standard
Dyson equation is employed

\begin{equation}
G(\omega)_{1p1h,1p'1h'}=G^{(2)}(\omega)_{1p1h}+G^{(2)}(\omega)_{1p1h} V_{1p1h,1p'1h'}
G(\omega)_{1p1h,1p'1h'}
\label{eq:dys},
\end{equation}
where we have used $G^{(2)}(\omega)_{1p1h}$ defined as,
\begin{equation}
G^{(2)}(\omega)_{1p1h} =
\frac{1}{\omega -  E_{1p1h}^{(2)}+ i \eta}
\label{eq:green2}
\end{equation}
instead of the free polarization propagator
$G^{(0)}(\omega)_{ph}$. By successive
iterations Eq.~(\ref{eq:dys}) leads to a
series which can be summed up only in some few particular
cases: when exchange terms are neglected or when
$V_{1p1h,1p'1h'}$ is a contact or a separable interaction. In the case that
exchange terms are neglected the series is known as ring series
and the sum of it as ring function.
The puzzle is then, how to deal with exchange
terms. For a general force, exchange terms should
be evaluated numerically order by order and in practice, this can
be done up to second order in the residual
interaction. Third and higher order contributions,
imply a quite involve numerical task due to the
number of diagrams and the dimensionality of the
integrals required to evaluate each diagram.
However, exchange terms of a contact interaction
can be easily included by a re-definition of the
constants entering into the interaction. Using
these facts, in Ref.~\cite{kn:ba96} we tackle this
problem as follows: we re-write the residual
interaction by summing and subtracting a contact
force, $V^C_{1p1h,1p'1h'}$,
\begin{equation}
V_{1p1h,1p'1h'} = V^C_{1p1h,1p'1h'} + V^F_{1p1h,1p'1h'}
\label{eq:int1}
\end{equation}
where $V^F_{1p1h,1p'1h'} \equiv V_{1p1h,1p'1h'} - V^C_{1p1h,1p'1h'}$
contains the finite
range part of the interaction. Varying $V^C_{1p1h,1p'1h'}$, the interaction
$V^F_{1p1h,1p'1h'}$ is adjusted in order to ensure a fast
convergence of the exchange terms. More
specifically, the different constant terms of
$V^F_{1p1h,1p'1h'}$ (that is, the ones of
$V^C_{1p1h,1p'1h'}$\footnote{Eventually, $V_{1p1h,1p'1h'}$ can contain
constant terms, but these terms can not be
changed. Only the ones of $V^C_{1p1h,1p'1h'}$ can be varied, as
this interaction is an artificial device to
guarantee the fast convergence of exchange terms
evaluated with $V^F_{1p1h,1p'1h'}$}), are varied until the
second order exchange contributions are almost
negligible compared with the first order ones.

The polarization propagator of Eq.~(\ref{eq:dys}) can now be
drawn as
\begin{equation}
G(\omega)_{1p1h,1p'1h'}=G(\omega)^{C}_{1p1h}+G(\omega)^{F}_{1p1h,1p'1h'}
+G(\omega)^{CF}_{1p1h,1p'1h'}
\label{eq:gw4},
\end{equation}
where,
\begin{eqnarray}
G^C_{1p1h} & = & G^{(2)}_{1p1h}+G^{(2)}_{1p1h} V^C_{1p1h,1p1h} G^{(2)}_{1p1h}+
G^{(2)}_{1p1h} V^C_{1p1h,1p1h} G^{(2)}_{1p1h} V^C_{1p1h,1p1h} G^{(2)}_{1p1h} + ... \nonumber \\
\label{eq:gw5a} \\ [9. mm]
G^F_{1p1h,1p'1h'} & = & G^{(2)}_{1p1h} V^F_{1p1h,1p'1h'} G^{(2)}_{1p'1h'}+
G^{(2)}_{1p1h} V^F_{1p1h,1p''1h''} G^{(2)}_{1p''1h''} V^F_{1p''1h'',1p'1h'} G^{(2)}_{1p'1h'}.
\label{eq:gw5b} \\ [9. mm]
G^{CF}_{1p1h,1p'1h'} & = & G^{(2)}_{1p1h} V^{C}_{1p1h,1p''1h''} G^{(2)}_{1p''1h''}
 V^{F}_{1p''1h'',1p'1h'} G^{(2)}_{1p'1h'}  + \nonumber \\
& + & G^{(2)}_{1p1h} V^{F}_{1p1h,1p''1h''} G^{(2)}_{1p''1h''}
V^{C}_{1p''1h'',1p'1h'} G^{(2)}_{1p'1h'}  + \nonumber \\
& + & G^{(2)}_{1p1h} V^{C}_{1p1h,1p'''1h'''} G^{(2)}_{1p'''1h'''}
V^{C}_{1p'''1h''',1p''1h''} G^{(2)}_{1p''1h''}
V^{F}_{1p''1h'',1p'1h'} G^{(2)}_{1p'1h'}+ ...   \
\label{eq:gw5c} \nonumber \\
\end{eqnarray}
Inserting  Eq.~(\ref{eq:gw4}) into  Eq.~(\ref{eq:rph2}) one can define
three different contributions to the response
function, $R^{C}_{ph}$, $R^{F}_{ph}$ and
$R^{CF}_{ph}$, associated to $G^{C}_{1p1h}$,
$G^{F}_{1p1h,1p'1h'}$ and $G^{CF}_{1p1h,1p'1h'}$,
respectively. In the case of $R^{C}_{ph}$, since
$V^C_{1p1h,1p'1h'}$ is a pure contact interaction, the direct
and exchange terms are equivalent and they can
both be summed up to infinite order by evaluating
the ring series of Eq.~(\ref{eq:gw5a}). In the case of
$R^{F}_{ph}$, Eq.~(\ref{eq:gw5b}) is only considered up to
second order in $V^F_{1p1h,1p'1h'}$, while for $R^{CF}_{ph}$,
Eq.~(\ref{eq:gw5c}) is also evaluated up to infinite order in
the $V^C_{1p1h,1p'1h'}$ interaction keeping terms up to second
order in $V^F_{1p1h,1p'1h'}$. Our final result is then,
\begin{equation}
R_{ph} = R^{C}_{ph} + R^{F}_{ph} + R^{CF}_{ph}
\label{eq:restot}
\end{equation}
Explicit expressions for the different exchange
terms needed to build up the $R^{F}_{ph}$ contribution
are very similar to the corresponding ones in the
transverse channels reported in Ref.~\cite{kn:ba96}.
There are minor differences in the multiplying constants,
the spin and isospin sums and the energy $E_{1p1h}^{(0)}$
should be replaced by $E_{1p1h}^{(2)}$. For this reason,
we do not reproduce them. We
neither show the ring series nor the direct terms as they
can be found in many references (see for example
Ref.~\cite{kn:fe71}).
If in Eq.~(\ref{eq:restot}) we replace
$G^{(2)}(\omega)_{1p1h}$ by the free propagator
$G^{(0)}(\omega)_{1p1h}$, the RPA approximation is
obtained. The first order direct contribution to the
RPA is drawn as the fourth diagram of the first line in
Fig.~1, while the next diagram is the corresponding exchange
term. Our approximation, where self-energies
are included in $G^{(2)}(\omega)_{ph}$, is usually
known as Second RPA (SRPA).

\subsection{$R_{\Delta h}$ contribution}

At variance with $R_{ph}$, the particular
character of ${\cal O}_{\Delta h}$ does not allow a
$1 \Delta 1h$ bubble to propagate as a RPA-series.
In this case, only the self-energy modifies the free
response. The expression for $R_{\Delta h}$ is obtained following
closely the steps done in the last sub-section to arrive to
$R_{ph}^{(2)}$. Obviously, one should replace $P_N$ by $P_{\Delta}$
and do some others minor changes. Before presenting the expression
for $R_{\Delta h}^{(2)}$, we would like to discuss briefly the
lower order contributions to the
$R_{\Delta h}$ response shown in Fig.~2: the first diagram in the {\it r.h.s.}
of the first line is the free response. The second and third
diagrams represent the Hartree-Fock self-energy over a hole
line. There is no Hartree term for the $\Delta$ due to the
absence of a scalar-isoscalar term in the $ V_{1 \Delta 1h,1 \Delta' 1h'}$
interaction. The fourth diagram in the first line of this figure is
the Fock contribution to the $\Delta$. Finally, in the second
line, we show the diagrams with self-energy insertions
$\Sigma^{P_{\Delta} Q P_{\Delta}}_{1 \Delta 1h}$.

Now we briefly outline the main expressions required to
obtain $R_{\Delta h}^{(2)}$. In Eq.~(\ref{eq:rnd})
we substitute $P_{\Delta}$ by it explicit expression,
\begin{eqnarray}
R_{\Delta h} & = & -\frac{1}{\pi} \ Im \sum_{1\Delta 1h, \, 1\Delta'1h'}
<|{\cal O}_{\Delta h}^{\dag}|1\Delta 1h><1\Delta 1h|
\frac{1}{\omega - H_{P_{\Delta} P_{\Delta}} - \Sigma^{P_{\Delta} Q P_{\Delta}} + i \eta}
 |1\Delta'1h'> \nonumber \\
 & & \times <1\Delta'1h'| {\cal O}_{\Delta h}|>
\label{eq:rdh}
\end{eqnarray}
The energy denominator of Eq. (\ref{eq:rdh}) is approximated as,
\begin{equation}
<1 \Delta 1h| H_{P_{\Delta} P_{\Delta}}  + \Sigma^{P_{\Delta} Q P_{\Delta}}
 |1 \Delta'1h'> \approx \delta_{\Delta h,\Delta'h'}
(E_{1 \Delta 1h}^{(0)} + \Sigma_{1 \Delta 1h}^{HF} +
\Sigma^{P_{\Delta} Q P_{\Delta}}_{1 \Delta 1h})
\label{eq:edd}.
\end{equation}
It is convenient to define the energies,
\begin{eqnarray}
E_{1 \Delta 1h}^{(0)} & = & \frac{p^2}{2 m_{\Delta}} -  \frac{h^2}{2 m}
+  m_{\Delta} - m, \nonumber \\
E_{1 \Delta 1h}^{(1)} & = & E_{1 \Delta 1h}^{(0)} + \Sigma^{F}_{\Delta}(p) - \Sigma^{HF}(h)
\nonumber \\
E_{1 \Delta 1h}^{(2)} & = & E_{1 \Delta 1h}^{(1)} +
\Sigma^{P_{\Delta} Q P_{\Delta}}_{1 \Delta 1h}(\v{q},\omega)
\label{eq:en012}
\end{eqnarray}
where expression for $\Sigma^{F}_{\Delta}(p)$
is given in Appendix B. We have adopted the same approximations
for $\Sigma^{P_{\Delta} Q P_{\Delta}}_{1 \Delta 1h}$
as for $\Sigma^{P_N Q P_N}_{1p1h}$. Explicit expressions for both
self-energies can be found in Ref.~\cite{kn:ba98}.

From Eq. (\ref{eq:rdh}) we define now,
\begin{equation}
R_{\Delta h}^{(i)}  =  -\frac{1}{\pi} \ Im \sum_{1 \Delta 1h}
|<1 \Delta 1h| {\cal O}_{\Delta h}|>|^2 \;
\frac{1}{\omega - E_{1 \Delta 1h}^{(i)} + i \eta}
\label{eq:rdh2}
\end{equation}
After some algebra it is easy to find,
\begin{equation}
R_{\Delta h}^{(0)}(\v{q},\omega) = \frac{8 \pi^2 A \, k_F}{m^4} \;
(G_{\Delta}(\v{q},\omega))^2 \;
(1 + \tau')
\; {\cal L}_{\Delta}(\v{q},\omega,m_{\Delta},m_h)
\label{eq:rdh0l}
\end{equation}

\begin{equation}
R_{\Delta h}^{(1)}(\v{q},\omega) = \frac{8 \pi^2 A \, k_F}{m^4}
\; (G_{\Delta}(\v{q},\omega))^2 \;
(1 + \tau')
\; {\cal L}_{\Delta}(\v{q},{\bar \omega},m_{\Delta},m_h)
\label{eq:rdhlm}
\end{equation}

\begin{equation}
R_{\Delta h}^{(2)}(\v{q},\omega) = - \, \frac{1}{\pi} \ \int_{0}^{\infty} dE
\; R_{\Delta h}^{(1)}(\v{q},E)
\frac{Im \Sigma^{P_{\Delta} Q P_{\Delta}}_{1 \Delta 1h}(\v{q},\omega)}{(E - \omega +
Re \Sigma^{P_{\Delta} Q P_{\Delta}}_{1 \Delta 1h}(\v{q},\omega))^2 +
(Im \Sigma^{P_{\Delta} Q P_{\Delta}}_{1 \Delta 1h}(\v{q},\omega))^2 } \
\label{eq:rdhse2},
\end{equation}
where in analogy with the $ph$-case, we have employed a modified
function ${\cal L}_{\Delta}(\v{q},\omega,m_{\Delta},m_h)$, which is
defined in Appendix A.

\newpage

\section{RESULTS}
In this section we show explicit results for the nuclear quasi-elastic
longitudinal response.
The properties of medium mass nuclei in the energy-momentum region of
interest are reasonably described by non-relativistic nuclear matter
once a proper Fermi momentum is used \cite{kn:am94}. In particular,
we consider the response of $^{40}Ca$ using a Fermi momentum
$k_F=235$ MeV/c.

For the residual interaction $V_{1p1h,1p'1h'}$, we have made use of
the parameterization of
the Bonn potential~\cite{kn:ma87},
as presented in Ref.~\cite{kn:br96}: it is expressed as
the exchange of $\pi$, $\rho$, $\sigma$ and $\omega$ mesons,
while the $\eta$ and $\delta$-mesons are neglected.
Also from this reference, we have taken the expressions for the
Hartree-Fock self-energy $\Sigma_{1p1h}^{HF}$.
When at least one $\Delta$ is involved, we have adopted the
model interaction already used in Ref. \cite{kn:ba98}, which for convenience is
reproduced in Appendix B. Employing these interactions, the first
step is to solve Eq.~(\ref{eq:rph2}) (and the corresponding one in
the $\Delta$ sector, $R_{\Delta h}^{(1)}$), in order to fix the
effective masses and energy shifts for particles, holes and
deltas. The results for several momentum transfers is shown in
Table~I.

Our main concern is to understand the longitudinal response and in
particular the effect of the $\Delta$ over it. As already stated
in the last section, the $\Delta$ affects the longitudinal
response via two mechanisms: by the direct excitation of a $\Delta
h$-pair and through the self-energies $\Sigma^{P_N Q_i P_N}_{1p
1h}$ and $\Sigma^{P_{\Delta} Q_i P_{\Delta}}_{1 \Delta 1h}$ with
$i=$ 2, 3 and 4. The first effect is considered in the $R_{\Delta
h}$ response function. In Fig.~3 we present results for
$R^{(0)}_{\Delta h}$, $R^{(1)}_{\Delta h}$ and $R^{(2)}_{\Delta
h}$  at momentum transfer $q=400$ MeV/c. The comparison of
$R^{(1)}_{\Delta h}$ with $R^{(0)}_{\Delta h}$ shows that the
effect of the Hartree-Fock self-energies is not relevant. At
variance, the influence of the $\Sigma^{P_{\Delta} Q
P_{\Delta}}_{1 \Delta 1h}$ is very significant: it produces an
important redistribution of the intensity to lower energies. The
$\Sigma^{P_{\Delta} Q P_{\Delta}}_{1 \Delta 1h}$ self-energy has
a real and an imaginary part and both terms are connected through
a dispersion relation. The effect of the imaginary part is to
spread the response, while the real part shifts the energy
position of the peak to lower values. These effects are very
strong because $\Sigma^{P_{\Delta} Q P_{\Delta}}_{1 \Delta 1h}$
increases with energy and the $\Delta h$ peak is high in energy.
It will be shown soon that
when we compare the magnitude of $R_{\Delta h}$ with $R_{ph}$
and with data, the $R_{\Delta h}$ response is as a whole negligible.
However, $R_{\Delta h}$ had been discussed for
completeness.

In Fig.~4 we analyze the effect of the different
contributions needed to built up the $R_{ph}$ response. In each panel, we have
drawn the free response, $R^{(0)}_{ph}$, the total response
$R_{ph}$, and the experimental points. This was done as a guidance
to understand the behaviour of each term separately. Let us resume
the meaning of dashed lines for each panel:

{\it Panel (a):} dashed lines represents the $R^{(1)}_{ph}$
response. Which means to study the effect of the Hartree-Fock
self-energy.

{\it Panel (b):} we show the RPA response; that is, in Eq.~(\ref{eq:restot})
we have employed $G^{(0)}(\omega)_{ph}$ instead of
$G^{(2)}(\omega)_{ph}$.

{\it Panel (c):} here we isolate the effect of the $\Sigma^{P_N Q_1 P_N}_{1p
1h}$ self-energy. In this panel,
we have evaluated Eq.~(\ref{eq:rphse2}) using a
re-defined energy, $\tilde{E}_{1p1h}^{(2)} =  \, E_{1p1h}^{(0)} \, + \,
\Sigma^{P_N Q_1 P_N}_{1p1h}(\v{q},\omega)$.

{\it Panel (d):} we evaluate Eq.~(\ref{eq:restot}) excluding the
$\Delta$ terms of the self-energy $\Sigma^{P_N Q P_N}_{1p1h}$
(which means to use $\Sigma^{P_N Q_1 P_N}_{1p1h}$ alone).

The Hartree-Fock self-energy $\Sigma_{1p1h}^{HF}$ spread the free
response and moves the quasi-elastic peak to higher energies, as
shown in {\it Panel (a)}. This result was already reported in
others works (see for example Ref. \cite{kn:br96}). The effect of
the RPA (see {\it Panel (b)}), depends on the character of the
interaction. In the longitudinal channel, the direct contribution
to the RPA (ring series), selects the isospin scalar end isospin
vector central terms of the interaction (usually named as $F$ and
$F'$). For our election of the interaction, these terms are
repulsive, which is reflected in the response. Turning now to
{\it Panel (c)}, we see that the action of the
$\Sigma^{P_N Q_1 P_N}_{1p1h}$ self-energy is similar
to the one in the $\Delta$-sector. However, the
magnitude of the spread and energy shift are smaller. Perhaps, the
most interesting result of this figure is the one in {\it Panel
(d)}. The comparison of our final result with the dashed line,
indicates that the $\Delta$ is very important in the longitudinal
channel, not because of it direct excitation, but due to its
indirect effect through the self-energy.

In Fig.~5 we compare our final result with data for several
momentum transfer $q$, ranging from 300 up to 500~MeV/c.
We have obtained a good
agreement at low values of $q$. Although the agreement for $q=$~450 and 500~MeV/c is
not so good, the accordance with the total intensity
is satisfactory.
At this point it is important to mention that the values of the
parameters entering into the residual interaction when the $\Delta$
is involved are quite uncertain.
This problem was already discussed in Ref.~\cite{kn:ba98} where we showed that
self-energy contributions which contains the $\Delta$,
change considerably due to this ambiguity.
Eventually, and in view of the result of {\it Panel (d)} in
Fig.~4, a more accurate interaction could improve the
agreement for high values of $q$. In addition, the
$ph$-interaction itself is not unambiguous. Let us quote Ref.~\cite{kn:br96}
where it has been pointed out that the bare nucleon-nucleon interaction
employed is too repulsive to reproduce the correct binding
energy and this shortcoming might also affect the RPA spectrum
(overestimating the hardening of the response). In summary, although
our results are encouraging and in fact we have obtained
perhaps one of the best agreements for the longitudinal response in a wide region
of $q$; these results are conditioned by the election of the
residual interaction which is yet unknown.

Concerning a comparison with other approaches, it is interesting
to comment on two works already mentioned in the Introduction,
where the longitudinal response with the inclusion of the $\Delta$
is considered. The first one is due to Gil {\it et al.}
(Ref.~\cite{kn:gi97}), where both the longitudinal and transverse
responses are evaluated. This work is very complex in the sense
that most of the many-body mechanisms where incorporated: mesons
exchange currens, polarization, initial and final state
interactions and the $\Delta$. The comparison with data is done
only at two momentum transfers: $q=$ 300 MeV/c for $^{12}C$ and
$q=$ 410 MeV/c for $^{40}Ca$. The last results are shown for
energies ranging between 40 and 220 MeV. Even if there is a small
tendency to overestimate data, these results certainly represents
an improvement over previous ones. Unfortunately, final values for
$^{40}Ca$ are shown for only one momentum transfer and in a narrow
energy region. In addition, the role of the $\Delta$ in the
longitudinal channel is not discussed in particular.
The second work was done by Amore {\it et al.} (Ref.~\cite{kn:am99b}),
where they have studied the effect of the $\Delta$ over the longitudinal
response using the BLE. As in our case, they have considered both the direct
excitation of a $\Delta h$-pair and the effect of the $\Delta$
over the response through many-body contributions.
They have also incorporated relativistic effects
and done the comparison with $^{12}C$ data for $q=$ 400, 500 and 600 MeV/c.
Their final results underestimate the experimental points.
Using a different formalism and interaction, this work
agrees with our results in two main issues:
the direct excitation of a $\Delta h$-pair is not
significant and the many-body effect of the $\Delta$ is important.

The main difference between our scheme and those of
Refs.~\cite{kn:gi97} and \cite{kn:am99b} is the way
in which we deal with the Pauli exchange terms.
In several works (see for instance
Refs.~\cite{kn:ba98}, \cite{kn:ba99}), we have determined that exchange terms are
important and need to be accurately evaluated. If we now incorporate
the transverse channel into the discussion, there are many
contributions like meson exchange currents and ground state
correlations beyond the RPA, which also need
to be considered. However, in the longitudinal channel these terms
are less relevant than in the transverse one.
Moreover, the interaction utilized in
Refs.~\cite{kn:gi97} and \cite{kn:am99b} is
different than ours (in fact, each work employs a different one).
This point is particularly important not only because of the
sensibility of the results over the interaction, but also because
antisymmetrization establishes constrains over the interaction (see
Ref.~\cite{kn:fr79}). Let us be more specific about these
constrains. When an antisymmetric matrix element is evaluated,
the antisymmetrization operator can act either over the initial
and final state or over the interaction itself. It is possible to
build up an 'antisymmetric' interaction where the prescription is
to evaluate direct matrix elements with this force. This is
equivalent to employ the non-antisymetrized interaction with
direct and exchange matrix elements.
When the antisymmetric interaction is employed, the different
channels of the interaction (for instance, $F$, $F'$, $G$ and $G'$ in a
Landau-Migdal type interaction), are not independent: a change in
one affects all the others.
As the RPA longitudinal and transverse
responses are dominated by different channels of the interaction,
the just mentioned constrains should be carefully considered when dealing with both
channels. Alternatively, the employment of any non-antisymetrized interaction
within a scheme which accurately evaluates exchange terms is
equivalent, as just mentioned.
In a forthcoming work we will discuss this
issue in detail~\cite{kn:ba03}.

\newpage

\section{CONCLUSIONS}
In the present work we have analyzed the longitudinal response
of the inclusive electron scattering process.
We have presented a formalism developed for
non-relativistic nuclear matter, which includes
correlations of the SRPA type. The $\Delta$-degree
of freedom was incorporated via two mechanisms:
the direct excitation of a $\Delta h$-pair and
through self-energy contributions which contains the
$\Delta$. Our scheme puts
special emphasis on the evaluation of Pauli
exchange terms.

By the employment of an effective Fermi momentum
we have evaluated the response for $^{40}Ca$ at
several momentum transfers, obtaining a good
agreement with data. These results were obtained
as a consequence of the addition of several
ingredients, which basically can be grouped into two
sets: the opening of new decay channels (as
$2p2h$, $1p 1 \Delta 2h$, etc.; represented by the
second order self-energies) which reduce the
height of the quasi-elastic peak and spread the
response, but moves the peak towards lower
energies. And in the second group we put the Hartre-Fock
self-energy and the RPA, which moves the peak to
higher energies, but do not provide the required
reduction of the intensity. In both cases the
action of the exchange terms and the election of
the interaction is very important.

In many works, it is mentioned that the
longitudinal response seems more elusive than the
transverse one, even though the underlying physics
is more complex in the last case. In view of our results, we
think that the just mentioned assertion should be
revised.
Referring this affirmation we recall that,  $i)$ the
first experimental results which separates the
longitudinal and transverse channels~\cite{kn:me84},
underestimate the longitudinal
response. $ii)$ the free Fermi response strongly
differs with data only in the longitudinal case.
$iii)$ A very simple model, which uses the ring
approximation with a $\pi$~+~$\rho$ plus the
Landau-Migdal $g'$-interaction gives a good
adjustment for the position and height of the
transverse peak. The conjunction of these elements
certainly shows the transverse channel as more
understandable. However, a word of caution
should be given: this simple model provides no
intensity in the dip region and when all possible
mechanisms (mesons exchange currents, initial and
final state interactions, etc.) are incorporated,
the resulting response shows a tendency to
overestimate data. Our model is more complex than the ring series
with the $g'$+$\pi$~+~$\rho$-interaction. But it is
simpler than the more elaborate ones for the
transverse channel already mentioned. These
points suggest that it is
precisely the transverse channel the most
difficult to be understood.

As a final comment, we would like to stress that
the many-body problem in inclusive quasi-elastic
electron scattering is quite involved. Any result
is strongly dependent on the nuclear interaction,
which is full of unknown coupling constants and
parameters. Recent improvements (see for instance,
Ref.~\cite{kn:gi97}), show that we are closer to the
solution of the puzzle, which necessarily implies a complex
many-body calculation together with an adequate election of the
nuclear interaction. In any case, the simultaneous study of
several momentum transfers should be given. In this sense, we think that the
present contribution is a step further in this direction. Clearly,
our next objective is the inclusion of ground state correlations and
meson exchange currents in order to analyze the transverse
channel.


\newpage

\section*{APPENDIX A}
In this appendix we give the explicit expression for ${\cal
L}(\v{q},\omega,m_p,m_h)$ and ${\cal L}_{\Delta}(\v{q},\omega,m_{\Delta},m_h)$
It is convenient to change variables using
the dimensionless quantities $\v{Q=q}/k_F$, $\v{h=h}/k_F$,
$\nu=\omega/ 2 \varepsilon_F$,
$\lambda_p=m/m_p$, $\lambda_h=m/m_h$, $\lambda_\Delta=m/m_\Delta$,
$c=\frac{m \, \lambda_\Delta}{m_\Delta \, \lambda_h}$ and
$\delta=\frac{m^2}{k_F^2} \frac{\lambda_h-\lambda_\Delta}{\lambda_h^2 \lambda_\Delta}$;
$\varepsilon_F=k_F^2/(2 m)$ is the Fermi
energy and $m$ is the nucleon mass.
\begin{equation}
{\cal L}(\v{Q},\nu,\lambda_p,\lambda_h) = m k_F  \int \frac{d^3 h}{(2 \pi)^3}
\theta(|\v{h}+\v{Q}| - 1) \theta(1 - |\v{h}|)
\delta (\nu -  ( \lambda_p (\v{h} + \v{q})^2 - \lambda_h \v{h}^2) )
\label{eq:lindm}
\end{equation}

If $\lambda_p=\lambda_h=1$,
\begin{equation}
{\cal L}(\v{Q},\nu,1,1) = \frac{m k_F}{(2 \pi)^2} \, \frac{1}{Q}
\, \nu
\label{eq:lind0}
\end{equation}
for $Q \leq 2$ and $0 \leq \nu \leq Q - \frac{1}{2} Q^2$,
\begin{equation}
{\cal L}(\v{Q},\nu,1,1) = \frac{m k_F}{(2 \pi)^2} \, \frac{1}{2 \, Q}
\, (1-(\frac{\nu}{Q}-\frac{Q}{2})^2)
\label{eq:lind02}
\end{equation}
for $Q \leq 2$ and $Q - \frac{1}{2} Q^2 \leq \nu \leq Q + \frac{1}{2}
Q^2$ or $Q \geq 2$ and $-Q + \frac{1}{2} Q^2 \leq \nu \leq Q + \frac{1}{2}
Q^2$ and ${\cal L}(\v{Q},\nu,1,1) = 0$ otherwise.

If $\lambda_p \neq \lambda_h$,
\begin{equation}
{\cal L}(\v{Q},\nu,\lambda_p,\lambda_h) = 0
\label{eq:lindm0}
\end{equation}
when $|\lambda_h - \lambda_p +2 \nu - \lambda_p Q^2|/(2 \, \lambda_p Q) >1$
or $\nu < (\lambda_p - \lambda_h)/2$,
\begin{equation}
{\cal L}(\v{Q},\nu,\lambda_p,\lambda_h) = \frac{m k_F}{(2 \pi)^2} \, \frac{1}{2 \, Q}
\, ( 2 \, \nu + (\lambda_h - \lambda_p))/(\lambda_h \lambda_p)
\label{eq:lindm1}
\end{equation}
for $Q \leq 2$ and $0 \leq \nu \leq (\lambda_p-\lambda_h (Q-1)^2)/2$,
\begin{eqnarray}
{\cal L}(\v{Q},\nu,\lambda_p,\lambda_h) & = & \frac{m k_F}{(2 \pi)^2} \, \frac{1}{2 \, Q}
\, [(\lambda_h - \lambda_p)^2-Q^2 \lambda_p (\lambda_h + \lambda_p)
+ 2 \nu (\lambda_h - \lambda_p) + \nonumber \\
& & + 2 Q \lambda_p
\sqrt{\lambda_h \lambda_p Q^2 - 2 \nu (\lambda_h - \lambda_p)}]
\frac{1}{\lambda_p (\lambda_h \lambda_p)^2}
\label{eq:lindm2}
\end{eqnarray}
for $Q \geq 2$ or $\nu \geq (\lambda_p-\lambda_h (Q-1)^2)/2$.

Now, we give the expression for
${\cal L}_{\Delta}(\v{Q},\nu,\lambda_{\Delta},\lambda_h)$,
\begin{equation}
{\cal L}_{\Delta}(\v{Q},\nu,\lambda_{\Delta},\lambda_h)= m k_F  \int \frac{d^3 h}{(2 \pi)^3}
\theta(1 - |\v{h}|) \, (h^2 Q^2 - (\v{h} \cdot \v{Q})^2) \,
\delta (\nu -  ( c (\v{h} + \v{q})^2/2 - \v{h}^2/2 + \delta) )
\label{eq:lindd}
\end{equation}
It is convenient to re-defined variables as,
\begin{eqnarray}
a_1 & = & \frac{1-c}{2} \nonumber \\
a_2 & = & -c |\v{Q}|  \nonumber \\
a_3 & = & \nu - \frac{c \, Q^2}{2} - \delta
\label{eq:nvar}
\end{eqnarray}
then, we get,
\begin{equation}
{\cal L}_{\Delta}(\v{Q},\nu,\lambda_{\Delta},\lambda_h)= 0
\label{eq:lindd2}
\end{equation}
if $|(a_1+a_3)/a_2| > 1$ and
\begin{eqnarray}
{\cal L}_{\Delta}(\v{Q},\nu,\lambda_{\Delta},\lambda_h) & = & \frac{2 \pi}{c^3 Q}
\; \{ - a_1^2/6+(a_2^2-2 a_1 a_3)/4 - c^2/2 + \nonumber \\
 & & + a_1^2 h_{min}^6/6-(a_2^2-2 a_1 a_3) h_{min}^4/4 + c^2 h_{min}^2/2\}
\label{eq:lindd3}
\end{eqnarray}
if $|(a_1+a_3)/a_2| \leq 1$, with $h_{min}=|(a_2 - \sqrt{a_2^2-4 a_1 a_3})/(2
a_1)|$.

\newpage

\section*{APPENDIX B}
In this appendix we show the residual interaction $V$ when at
least one $\Delta$ is involved and we present
the explicit expression for the Fock
self-energy for the $\Delta$, $\Sigma^{F}_{\Delta}(p)$.

We have employed the following $V_{1 \Delta 1h,1h' 1 \Delta'}$
interaction (this interaction is required in the evaluation of the
fourth diagram in the first line of Fig.~2 and in the second one
of the second line, already called as $Q_2$),
\begin{equation}
V_{1 \Delta 1h,1 \Delta' 1h'}(\v{p})=  \frac{f_{\pi \Delta N}^2}
{\mu_{\pi}^2} \,
\Gamma_{\pi \Delta N}^2 (p) \,
(\tilde{g'}_{\Delta N} (p)  \,
\v{\sigma} \cdot \v{S} +
\tilde{h'}_{\Delta N} (p) \, \v{\sigma} \cdot \widehat{\v{p}} \,
\v{S} \cdot \widehat{\v{p}}) \;
\v{\tau} \cdot  \v{T}
\label{eq:intnd}
\end{equation}
where,
\begin{eqnarray}
\tilde{g'}_{\Delta N} (p) & = & g_{\Delta N}' -
\frac{\Gamma_{\rho \Delta N}^2 (p)}
{\Gamma_{\pi \Delta N}^2 (p)}\ C_{\rho \Delta N}
\frac{p^2}{p^2 + \mu_{\rho}^2}, \nonumber \\
\tilde{h'}_{\Delta N} (p) & = &
-\frac{p^2}{p^2 + \mu_{\pi}^2} \ +
\frac{\Gamma_{\rho \Delta N}^2 (p)}
{\Gamma_{\pi \Delta N}^2 (p)}\ C_{\rho \Delta N}
\frac{p^2}{p^2 + \mu_{\rho}^2},
\label{eq:hpt}
\end{eqnarray}
and
\begin{equation}
\Gamma_{\pi \Delta N, \rho \Delta N} (p)  =
\frac{\Lambda_{\pi \Delta N, \rho \Delta N}^2 - (\mu_{\pi, \rho})^2 }
{\Lambda_{\pi, \rho}^2 + p^2} \ ,
\label{eq:fft}
\end{equation}
Analogous expressions are used for $V_{1 \Delta 1h,1p'1h'}$
(needed in the evaluation of the first diagram of the second line
in Fig.~2, called $Q_1$),  $V_{1 \Delta 1h, 1 \Delta' 1h'}$ ($Q_3$) and
$V_{1 \Delta 1h,2 \Delta}$ ($Q_4$). For each of these interactions, the coupling
constants and parameters should be replace
by their corresponding $NN$, $N \Delta$ and $\Delta \Delta$ values.
Also Pauli matrices must be replaced by the corresponding
transitions matrices $\v{S}$ and $\v{T}$
or $\v{{\cal S}}$ and $\v{{\cal T}}$,
the 3/2-3/2 spin matrices (see ref. \cite{kn:ce97}); depending on
the character of the mesonic vertex.

For the parameters entering into our theory we have set at
140 MeV (770 MeV) the mass of the pion (rho meson). The
pion coupling constant $f_{\pi NN}^2/4 \pi$=0.081,
$f_{\pi \Delta N}$=2 $f_{\pi NN}$ and
$f_{\pi \Delta \Delta}$= $\frac{4}{5} \, f_{\pi NN}$.
For the rho meson we have used
$C_{\rho NN} = C_{\rho \Delta N} = C_{\rho \Delta \Delta}$ = 2.18.
The different mesons cut-offs at the vertices are set to
$\Lambda_{\pi NN}$ = 1300 MeV/c, $\Lambda_{\rho NN}$ = 1750 MeV/c
while all the remaining cut-offs are set to 1000 MeV/c.
For the Landau Migdal parameter $g'_{\Delta N}$ and
$g'_{\Delta \Delta}$ we have taken
the values 0.55 and 0.40, respectively.

We present now the expression for $\Sigma^{F}_{\Delta}(p)$,
\begin{eqnarray}
\Sigma^{F}_{\Delta}(\v{p}) & = & - \frac{2 \pi k_F}{3} \frac{f_{\pi \Delta N}^2}{\mu_{\pi}^2}
\; \{(3 g'_{\Delta N}-1) \, v_0(\Lambda_{\pi \Delta N},\mu_{\pi},\v{p})
- 2 \,  C_{\rho \Delta N} \,  v_0(\Lambda_{\rho \Delta N},\mu_{\rho},\v{p})
\nonumber \\
& & + v_c(\Lambda_{\pi \Delta N},\mu_{\pi},\v{p})
+ 2  \, C_{\rho \Delta N} \,  v_c(\Lambda_{\rho \Delta N},\mu_{\rho},\v{p}) \}
\label{eq:fsed}
\end{eqnarray}
where we have introduced the functions $v_0$ and $v_c$, defined
as,
\begin{eqnarray}
v_0(\Lambda,\mu,\v{p}) & = & \int \frac{d^3 k}{(2 \pi)^3}
\; \theta(1-|\v{k}|) \;
(\frac{\Lambda^2-\mu^2}{\Lambda^2+(\v{k} -\v{p})^2})^2
\nonumber \\
v_c(\Lambda,\mu,\v{p}) & = & \int \frac{d^3 k}{(2 \pi)^3}
\; \theta(1-|\v{k}|) \;
\frac{\mu^2}{\mu^2+(\v{k} -\v{p})^2} \;
(\frac{\Lambda^2-\mu^2}{\Lambda^2+(\v{k} -\v{p})^2})^2
\label{eq:v0vc}
\end{eqnarray}
which can be easily integrated.

\newpage

\begin{quote}

\centering Table I.

\vspace{7 mm}

\begin{tabular}{cccccccc}   \hline\hline
$q$ [MeV/c] & \multicolumn{3}{c} {$R_{ph}^{(1)}$} & &
\multicolumn{3}{c} {$R_{\Delta h}^{(1)}$} \\
\cline{2-4} \cline{6-8}
  & $~~~\lambda_p$~~~ & $~~~\lambda_h$~~~ & $\Delta \omega$ [MeV]&
& ~~~$\lambda_{\Delta}$~~~ & ~~~$\lambda_h$~~~ & $\Delta \omega$ [MeV] \\ \hline
300.  & 1.33 & 1.54 & 11.0 & & 1.08 & 1.11 & ~4.0 \\
325.  & 1.32 & 1.54 & 11.0 & & 1.06 & 1.10 & ~3.5 \\
350.  & 1.31 & 1.54 & ~9.0 & & 1.05 & 1.09 & ~3.0 \\
400.  & 1.28 & 1.54 & ~9.0 & & 1.05 & 1.08 & ~3.0 \\
450.  & 1.32 & 1.56 & ~9.0 & & 1.04 & 1.07 & ~2.5 \\
500.  & 1.37 & 1.59 & ~7.0 & & 1.04 & 1.07 & ~2.0 \\  \hline\hline \\
\end{tabular}

\end{quote}

\vspace{15 mm}

Table I: Effective masses and energies shifts employed in
$R_{ph}^{(1)}(\v{q},\omega)$ and $R_{\Delta h}^{(1)}(\v{q},\omega)$.
We have used dimensionless quantities $\lambda_{m_i} \equiv
m/m_i$, where $i$ = $p$, $h$ or $\Delta$.

\newpage

\vfill
\eject
\section*{Figure Captions}
\begin{description}

\item [Figure 1:]
Goldstone diagrams stemming from Eq.~(\ref{eq:restot}).
In every diagram two wavy lines represent the external
probe with momentum and energy $\v{q}$ and $\omega$,
respectively. The dashed line represents the residual interaction.
Continuous lines represent either a particle or a hole, while a
double-continuous line represents the $\Delta$.
Only forward-going contributions are shown.

\item [Figure 2:]
The same as Fig.~1, but for Eq.~(\ref{eq:rdhse2}).

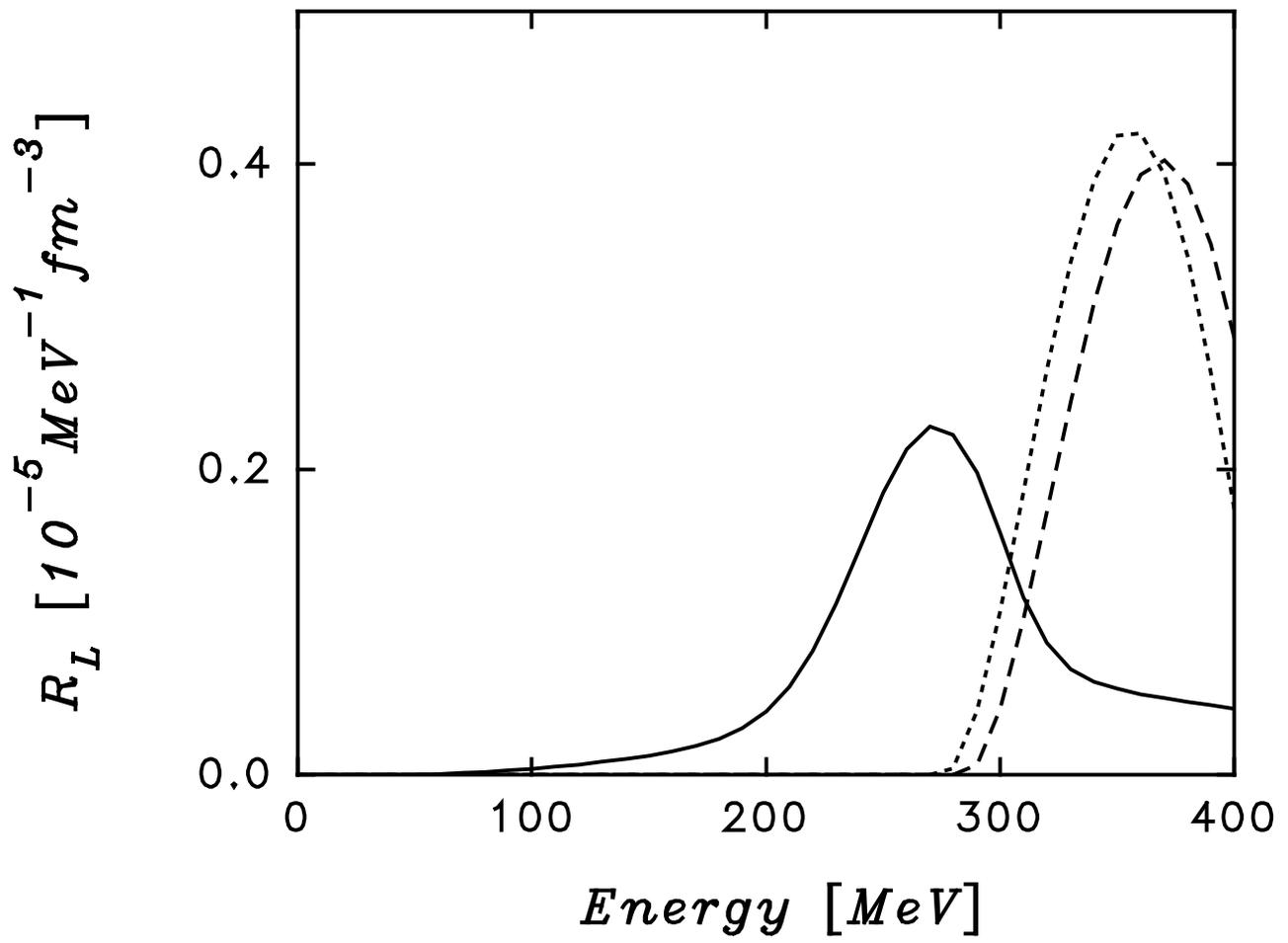
\item [Figure 3:]
The $R_{\Delta h}$ response at momentum transfer $q=$ 400 MeV/c.
The short-dashed line represents the
free response of Eq.~(\ref{eq:rdh0l}), the long-dashed line the
$R_{\Delta h}^{(1)}$ response of Eq.~(\ref{eq:rdhlm}), while the
continuous line is the final result $R_{\Delta h}^{(2)}$ of
Eq.~(\ref{eq:rdhse2}).

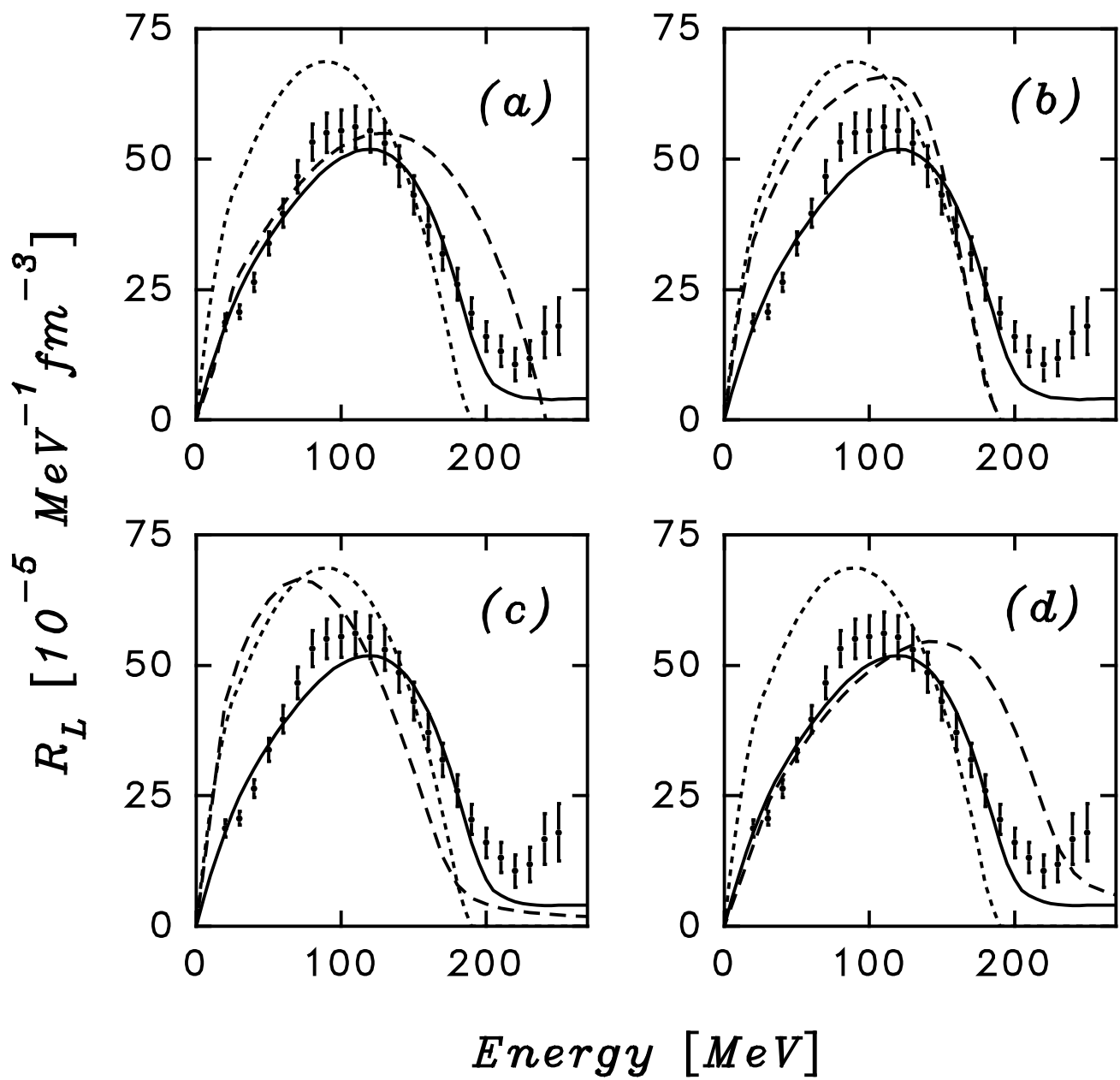
\item [Figure 4:]
Different contributions to the $R_{ph}$ response. In each panel
the short-dashed line represents the free response, the continuous
line is the final result of Eq.~(\ref{eq:restot}) and the meaning of
the long-dashed line for each panel is explained in the text. The
momentum transfer is 400 MeV/c for all panels. Data was taken
from Ref.~\cite{kn:wi97}.

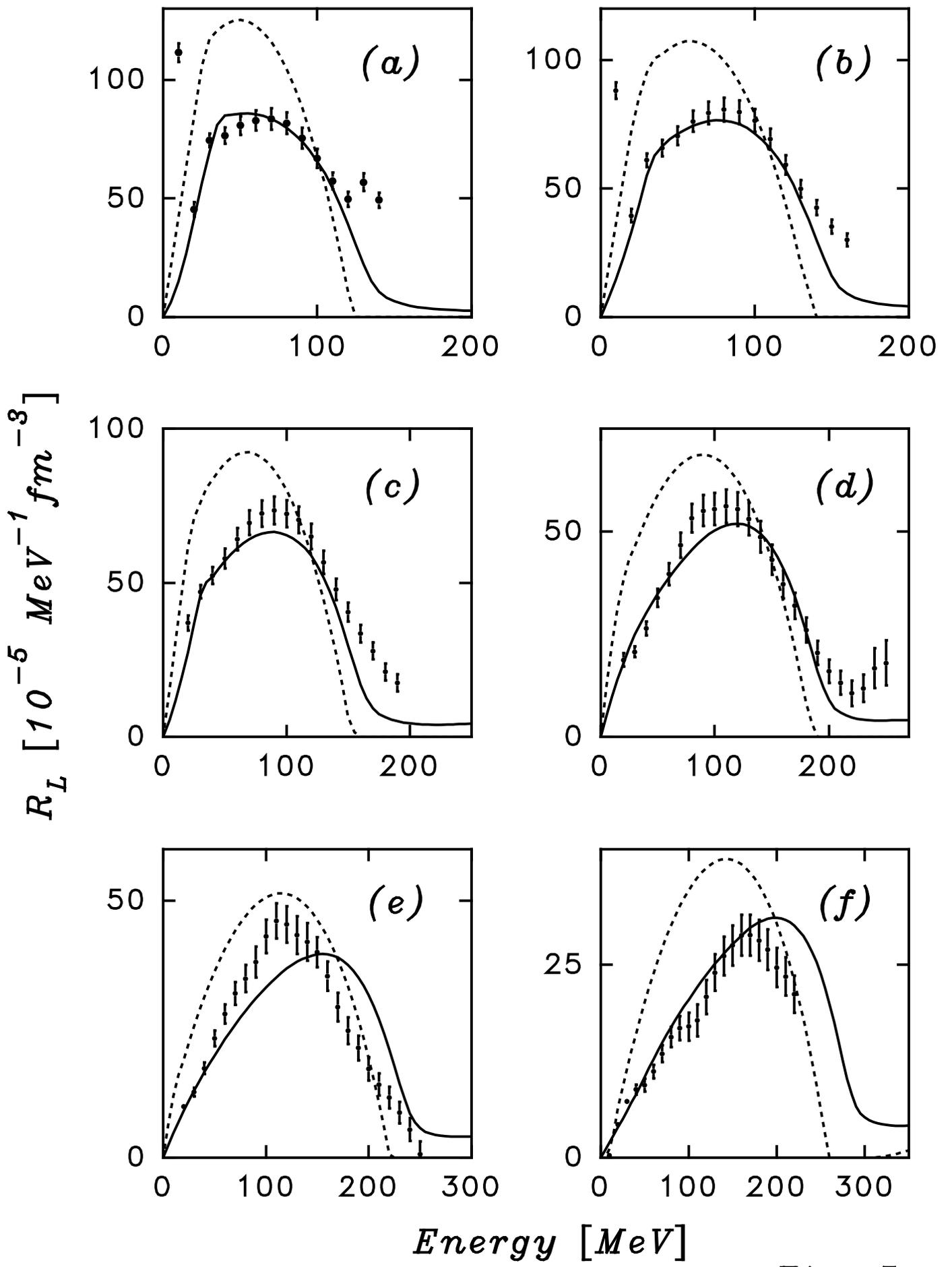
\item [Figure 5:]
The $R_{ph}$ response at different momentum transfer. Short-dashed
line are the free responses and the continuous ones are the sum of
$R_{ph}$ plus $R_{\Delta h}$. The momentum transfer in each panel
is: (a) 300 MeV/c, (b) 325 MeV/c, (c) 350 MeV/c, (d) 400 MeV/c,
(e) 450 MeV/c and (f) 500 MeV/c.  Data was taken
from Ref.~\cite{kn:wi97}.

\end{description}
\vfill
\eject

\end{document}